# Comment on " Formulas and numerical table for the radial part of overlap integrals with the same screening parametres of Slater-type orbitals "

I.I. Guseinov

*Department of Physics, Faculty of Arts and Sciences, Onsekiz Mart University, Çanakkale, Turkey*

**Abstract**

Recently published formulas for the calculation of radial part of overlap integrals (E.Öztekin, M.Yavuz, S.Atalay, Theor.Cmem.Acc., (2001), 106, 264) are critically analyzed. It is demonstrated that the presented in this work formulas are not original and they can be easily be obtained from the formulas already established in our papers.

Recently Öztekin, Yavuz and Atalay in Ref.[1] published the formulas for the radial part of overlap integrals with the same screening parametres of STOs. Most of these formulas are available in our published papers or can eaşily be derived by means of a simple algebra from the following equations of Refs.[2-4]:

Eqs.(19),(20) and (21) of Ref.[2] (compare with formulas of Section 2.1 in Ref.[1]):

$$g_{\alpha\beta}^{0}(l\lambda,l'\lambda) = \sum_{k=0}^{\lambda}(-1)^{k}F_{k}(\lambda)D_{\alpha+2\lambda-2k}^{l\lambda}D_{\beta}^{l'\lambda}$$

(1)

$$D_{\beta}^{l\lambda} = \frac{(-1)^{(1-\beta)/2}}{2^{l}}\left[\frac{2l+1}{2}\frac{F_{l}(l+\lambda)}{F_{\lambda}(l)}\right]^{1/2} F_{(1-\beta)/2}(l)F_{\beta-\lambda}(1+\beta) \qquad (2)$$

$$F_{m}(N,N') = \sum_{k}(-1)^{k}F_{m-k}(N)F_{k}(N') \qquad ; \qquad (3)$$

Eqs.(12), (13), (17), (19), (20), (21), (22), (24), (25) and (26) of Ref.[3] (compare with the formulas of Section 2.2 in Ref.[1]):

$$S_{nlm,n'l'm'}(\vec{p}) = \sum_{N=1}^{n+n'+1}\sum_{L=0}^{N-1}\sum_{M=-L}^{L} g_{nlm,n'l'm'}^{NLM}\chi_{NLM}^{*}(1,\vec{p}) \qquad (4)$$

$$g_{nlm,n'l'm'}^{NLM} = \sum_{N'=1}^{n+n'+1}\Omega_{NN'}^{L}(n+n'+1)T_{nlm,n'l'm'}^{NLM} \qquad (5)$$

$$T_{nlm,n'l'm'}^{NLM} = 2^{3/2}\pi(-1)^{(l-l'-L)}\sqrt{\frac{2L+1}{4\pi}}C^{LM}(lm,l'm')A_{mm'}^{M}Q_{nl,n'l'}^{NL} \qquad (6)$$

$$Q_{nl,n'l'}^{NL} = [F_{l}(n)F_{l'}(n')F_{L}(N)\sqrt{F_{n}(2n)F_{n'}(2n')F_{N}(2N)}]^{-1}$$

$$\times \sum_{s=0}^{E((n-1)/2)+E((n'-l'/2)+E((N-L)/2)} (-1)^s a_s(l+1, n-l; '+1, n'-l'; L+1, N-L) \tag{7}$$

$$\times \sum_{m=0}^{g+1} (-1)^{l+2g-2m} F_m(g+1) F_\gamma(2\gamma-1)$$

$$C_n^\alpha(x) = \sum_{s=0}^{E(n/2)} (-1)^s a_s(\alpha, n)(2x)^{n-2s} \tag{8}$$

$$C_n^\alpha(x) C_{n'}^{\alpha'}(x) = \sum_{s=0}^{E(n/2)+E(n'/2)} (-1)^s a_s(\alpha, n, \alpha', n')(2x)^{n+n'-2s} \tag{9}$$

$$C_n^\alpha(x) C_{n'}^{\alpha'}(x) C_{n''}^{\alpha''}(x) = \sum_{s=0}^{E(n/2)+E(n'/2)+E(n''/2)} (-1)^s a_s(\alpha, n, \alpha', n', \alpha'', n'')(2x)^{n+n'+n''-2s} \tag{10}$$

$$a_s(\alpha, n) = F_{\alpha-1}(\alpha-1+n-s) F_s(n-s) \tag{11}$$

$$a_s(\alpha, n, \alpha', n') = \sum_{m=0}^{E(n/2)} a_m(\alpha, n) a_{s-m}(\alpha', n') \tag{12}$$

$$a_s(\alpha, n, \alpha', n', \alpha'', n'') = \sum_{m=0}^{E(n/2)} a_m(\alpha, n) a_{s-m}(\alpha', n', \alpha'', n'') \tag{13}$$

Eqs.(4), (9), (11), (12), (13), (15) and (17) of Ref.[4] (compare with sentences and formulas of Section 2.3 and 3.3 in Ref.[1]):

$$S_{nl\lambda, n'l'\lambda'}(p, t) = \sum_{l''=\lambda}^{l} \frac{[2p(1+t)]^l}{[2p(1-t)]^{l''}} \left[ \frac{(2l+1)(2l'')! F_{2n'}(2n'+2l'') F_{l''+\lambda}(1+\lambda) F_{l''-\lambda}(1-\lambda)}{(2l''+1)(2l)! F_{2n-2l}(2n)} \right]^{1/2}$$

$$\times \sum_L \sqrt{2L+1} C^L(l'\lambda, l''\lambda) S_{n-l00, n'+l''L0}(p, t), \tag{14}$$

$$S_{n, n'l'}(p, t) = -A_{l'-1} \left\{ \frac{p(1-t)}{[(2n'-1)2n']^{1/2}} S_{n, n'-1l'-1}(p, t) + \frac{[(2n'+1)(2n'+2)]^{1/2}}{4p(1-t)} S_{n, n'+1l'-1}(p, t) \right.$$

$$\left. - \frac{(1-t)}{4p[(1+t)]^2} \left[ \frac{(2n+1)(2n+2)(2n+3)(n+2)}{(2n'-1)n'} \right]^{1/2} S_{n+2, n'-1l'-1}(p, t) \right\} - B_{l'-1} S_{n, n'l'-2}(p, t), \tag{15}$$

$$S_{0n'}(p, 0) = \left[ \frac{n}{2(2n'-1)} \right]^{1/2} S_{0n'-1}(p, 0) + \left[ \frac{2(2n'+1)}{n'+1} \right]^{1/2} \eta_{0n'+1}(p, 0) e^{-p} \tag{16}$$

$$S_{nn'}(p, 0) = \left[ \frac{n(2n'-1)}{(2n-1)(n'+1)} \right]^{1/2} S_{n-1n'+1}(p, 0) - \left[ \frac{n(n-1)(2n'+1)}{2(2n-1)(2n-3)(n'+1)} \right]^{1/2} S_{n-2n'+1}(p, 0)$$

$$+\left[\frac{n'}{2(2n'-1)}\right]^{1/2} S_{nm'-1}(p,0) \tag{17}$$

$$\eta_{nm'}(p,t) = \frac{[2p(1+t)]^{n+1/2}[2p(1-t)]^{n'+1/2}}{4p^2\left[(2n)!(2n')!\right]^{1/2}} \tag{18}$$

$$S_{00}(p,0) = e^{-p} \tag{19}$$

$$S_{nlm,n'l'm'}(p,t;\theta,\phi) = \frac{\sqrt{(2n+1)(2n+2)}}{2p(1+t)\cos\theta}\left[A_{l\lambda}S_{n+1l+1m,n'l'm'}(p,t;\theta,\phi) + B_{l\lambda}S_{n+1l-1m,n'l'm'}(p,t;\theta,\phi)\right]$$

$$-\frac{\sqrt{(2n'+1)(2n'+2)}}{2p(1-t)\cos\theta}\left[A_{l\lambda'}S_{nlm,n'+1l'+1m'}(p,t;\theta,\phi) + B_{l\lambda'}S_{nlm,n'+1l'-1m'}(p,t;\theta,\phi)\right] \tag{20}$$

As can be seen from these equations and the formulas Ref.[1] the authors have slightly modified the formulas presented in Ref.[2-4] (amazingly, several sentences were copied verbatim). We notice that, by means of formulas for overlap integrals over STOs obtained in our papers we have had considerable success in the calculation of multicenter integrals over STOs containing in the Hartree-Fock- Roothaan equations for molecules (see, e.g., Refs.[5]).